\documentclass{vgtc}                         

\ifpdf
  \pdfoutput=1\relax                   
  \usepackage{graphicx}                
  \DeclareGraphicsExtensions{.pdf,.png,.jpg,.jpeg} 
\else
  \usepackage{graphicx}                
  \DeclareGraphicsExtensions{.eps}     
\fi%

\graphicspath{{figures/}{pictures/}{images/}{./}} 

\usepackage{microtype}                 
\PassOptionsToPackage{warn}{textcomp}  
\usepackage{textcomp}                  
\usepackage{mathptmx}                  
\usepackage{times}                     
\usepackage{cite}                      
\usepackage{tabu}                      
\usepackage{booktabs}                  

\usepackage{enumitem}
\usepackage{xspace}
\usepackage[pscoord]{eso-pic}

\newcommand{\placetextbox}[3]{
  \setbox0=\hbox{#3}
  \AddToShipoutPictureFG*{
    \put(\LenToUnit{#1\paperwidth},\LenToUnit{#2\paperheight}){\vtop{{\null}\makebox[0pt][c]{#3}}}%
  }%
}%
\newcommand\ie{i.\,e.\xspace}
\newcommand\eg{e.\,g.\xspace}
\newcommand\cf{cf.\xspace}

\newcommand{\cue}[1]{\textsc{\lowercase{#1}}}

\usepackage{csquotes}
\usepackage{amsmath}
\usepackage{algorithmic}
\usepackage{graphicx}
\usepackage{subcaption}
\usepackage{stfloats}
\usepackage{threeparttable}
\usepackage{tabularx}

\vgtccategory{Research}

\title{Exploring the Effect of Visual Cues on Eye Gaze During AR-Guided Picking and Assembly Tasks}

\author{Paper ID:}
\author{Arne Seeliger\thanks{e-mail: aseeliger@ethz.ch}\\ %
        \scriptsize{ETH Zurich} %
\and Gerrit Merz\\ %
     \scriptsize{Karlsruhe Institute of Technology} %
\and Christian Holz\\ %
     \scriptsize{ETH Zurich} 
\and Stefan Feuerriegel\\ %
     \scriptsize{ETH Zurich}}

\abstract{In this paper, we present an analysis of eye gaze patterns pertaining to visual cues in augmented reality (AR) for head-mounted displays (HMDs). We conducted an experimental study involving a picking and assembly task, which was guided by different visual cues. We compare these visual cues along multiple dimensions (in-view vs. out-of-view, static vs. dynamic, sequential vs. simultaneous) and analyze quantitative metrics such as gaze distribution, gaze duration, and gaze path distance. Our results indicate that visual cues in AR significantly affect eye gaze patterns. Specifically, we show that the effect varies depending on the type of visual cue. We discuss these empirical results with respect to visual attention theory.%
} 

\CCScatlist{
  \CCScatTwelve{Human-centered computing}{Ubiquitous and mobile computing}{Empirical studies in ubiquitous and mobile computing}
}

\begin{document}
\maketitle
\section{Introduction}
\placetextbox{0.5}{0.05}{\fbox{\textsf{Paper accepted at the 2021 IEEE International Symposium on Mixed and Augmented Reality Adjunct (ISMAR-Adjunct) ©2021 IEEE}}}

Eye gaze analysis is frequently employed to better understand how users interact with computer interfaces. It has been extensively used in the context of two-dimensional interfaces like screens. Common applications are, for example, the analysis of eye gaze in web page interactions \cite{Georges2016UXInterfaces}. Eye gaze analysis has also been used in combination with subtle image modulations to direct visual attention \cite{Bailey2009SubtleDirection}. 

Typically, evidence is obtained through desktop-based experiments. However, when studying eye gaze using mobile or wearable devices \cite{Jiang2019ApplyingMechanisms}, a transfer of findings from desktop-based settings is difficult, especially when body movement and physical surroundings need to be accounted for \cite{Gullberg2002}. Moreover, eye gaze behavior might differ substantially between static and dynamic settings \cite{Kinsman2012Ego-motionTracking}.


Considering AR, specifically for HMDs, empirical assessment of eye gaze remains rare. Renner and Pfeiffer \cite{Renner2017AttentionSystems} compare AR-based attention guiding techniques in assembly tasks. However, they mainly employ evaluation metrics related to task performance (\eg, completion time). Similarly, Burova et al. \cite{Burova2020UtilizingMaintenance} examine AR-based guidance and safety awareness cues. Despite calculating fixation counts, the work is mainly based on self-assessments (\ie, questionnaires). Moreover, the above studies use virtual reality (VR) devices to simulate AR environments. Yet, it remains open whether such findings are transferable to AR settings.


In this paper, we provide an analysis of eye gaze patterns relating to AR-based guidance cues in picking and assembly tasks. In one user study, $12$ participants were guided by visual cues, which were displayed through an HMD. Specifically, eight visual cues were used, which differed along multiple dimensions (in-view vs. out-of-view, static vs. dynamic, sequential vs. simultaneous). We recorded eye gaze data for all participants and inferred quantitative metrics, including gaze distribution, gaze duration, and gaze path distance. Based on this, we discuss the empirical results in light of visual attention theory.

\section{Related Work}
\subsection{Visual Attention and Search}
At any given moment, the human visual system is exposed to more perceptual information than can be processed at the same time \cite{Chun2000BlackwellPerception}, which makes visual attention and search necessary \cite{Wolfe2020VisualFor}. Visual attention enables the active selection of relevant information and the ignoring of irrelevant information from a complex visual environment \cite{Chun2000BlackwellPerception}. Although attention and eye movement are closely entangled \cite{Wolfe2020VisualFor}, it is possible to fixate one location while attending another \cite{Kelley2008CorticalAttention}. Therefore, it is common to distinguish between overt eye movement and covert deployment of attention \cite{Posner1980OrientingAttention.}. Overt attention can be measured effectively by an eye tracker while the tracking of covert attention poses greater challenges \cite{Wolfe2020VisualFor}.

Visual search characterizes a situation in which a subject looks for a target item among multiple distractor items \cite{Wolfe1994GuidedSearch}. Many theories regarding the mechanisms of search and search efficiency have been proposed (\eg, Treisman’s feature integration theory (FIT) \cite{Treisman.A1980AAttention}). In this context, two forms of guidance are often distinguished, namely bottom-up, stimulus-driven attention and top-down, goal-driven attention \cite{Wolfe2020VisualFor}. In stimulus-driven attention, attention is attracted automatically through a salient stimulus, whereas goal-driven attention is under overt control of the human observer \cite{Chun2000BlackwellPerception}. 

\subsection{Attention Guidance in Augmented Reality}
\label{attention_guidance_AR}

Attention guidance towards areas-of-interest (AOIs) can be achieved by directing the user's attention through the use of visual cues that indicate the target location \cite{Hoffmann2008EvaluatingScreens}. 
Visual cues can either be presented at a fixed point on the display (in-view) or affixed to the focus object (in-situ) \cite{Renner2017AttentionSystems}. In this terminology, a conventional \cue{on-screen arrow} (\eg, \cite{Alghofaili2019LostNavigation}) is classified as in-view, whereas an arrow placed within the environment and pointing towards an object of interest is classified as in-situ. In case the object of interest lies outside a user's field-of-view (FOV), the visual cue first needs to guide the user's attention to the off-screen location. 

There are various examples of visual cues. \cue{Halo} is a well-known cue for visualizing off-screen locations \cite{Baudisch2003Halo:Objects}. For this, \cue{Halo} shows circles around off-screen objects, where the circles are sufficiently large to reach into the border region of the display window. However, \cue{Halo} has limited capacity when displaying locations of multiple objects in the same corner. This is addressed by \cue{Wedge} \cite{Gustafson2008Wedge:Locations}, where circles are replaced by isosceles triangles. The \cue{attention funnel} \cite{Biocca2006} uses a tunnel of frames that are drawn from the central FOV towards the location of the target object. Other cues include, for instance, \cue{Deadeye} \cite{Krekhov2019Deadeye:Presentation}. 
Note that most of the aforementioned cues were originally developed for two-dimensional interfaces such as screens but they have also been found suitable for HMDs \cite{Gruenefeld2017VisualizingReality}.

\subsection{Eye Gaze Metrics}
\label{gaze_metrics}
Common metrics for eye gaze analysis include the following. (1)~Eye fixations point to individual locations of user attention, thus yielding gaze distributions. A larger number of fixations can indicate higher importance or noticeability of an AOI \cite{Poole2005EyeProspects}. A larger total count of fixations has been associated with inefficient search since this indicates that many irrelevant elements have been sampled \cite{Goldberg1999ComputerConstructs}. 
(2)~Dwells provide an aggregated metric, often to describe gaze duration. For this, multiple consecutive fixations on the same AOI are counted as a single dwell \cite{Goldberg2002EyeImplications, Jiang2019ApplyingMechanisms, Gullberg2002}. The number of dwells and dwell times thus link to the overall attention per AOI. Dwell times may also be taken as an indicator of task difficulty \cite{Stork2010HumanApplications}.
(3)~The time to first fixation (TTFF) relates to how quickly an object has captured a user's attention. It is defined as the time between stimulus onset and its first fixation. Thus, TTFF can be used to assess an object's property to attract attention \cite{Byrne1999EyeMenus}. 
(4)~Scanpaths refer to the path defined by successive points-of regard (PORs) (\eg, \cite{Goldberg2002EyeImplications}). Scanpaths help to understand user search behavior by indicating more or less efficient search \cite{Goldberg1999ComputerConstructs}.
(5)~Saccades refer to rapid eye movements between fixations and are quantified typically according to  saccadic amplitudes and saccadic velocity  \cite{Goldberg2002EyeImplications, Goldberg1999ComputerConstructs}. Saccades relate to both cognitive and non-conscious processes and are thus relevant for studying affective information processing, which is outside of our study objective. 

\section{AR Guidance and Eye Tracking System}
To asses gaze patterns in AR, we developed a system consisting of: (1)~an HMD for showing visual cues, (2)~eye tracking for identifying PORs, and (3)~eye gaze analysis through quantitative metrics.

\subsection{Head-Mounted Display for Guiding Attention in AR}
Our system guides user attention by showing different AR-based visual cues through an HMD. Specifically, it is designed to run on HoloLens~2 and, for this, we used the Mixed Reality Toolkit and Unity. For augmented viewing, HoloLens~2 provides a horizontal FOV of 43\textdegree{} and a vertical FOV of 29\textdegree{}. However, no details about the FOV of its eye tracking system have been released so far. To give an approximation, we measured at which angles the eye tracking system of HoloLens~2 was able to record gaze points. We observed a horizontal FOV of approximately 40\textdegree{} in both directions and a vertical FOV of approximately 20\textdegree{} in the upper direction and 40\textdegree{} in the lower direction.

\subsection{Eye Tracking for Identifying PORs and Fixated AOIs}
For eye tracking, the system uses the HMD's built-in eye tracking sensors, which capture eye movement at a maximum sampling rate of 30\,Hz. HoloLens~2 provides gaze rays that lie within 1.5\textdegree{} of visual angle \cite{Microsoft2020Eye2}. Since manufacturer specifications can be inaccurate in real-world scenarios \cite{Feit2017TowardDesign}, we validated the hardware by examining accuracy and precision at three distances (0.5\,m, 1\,m, and 2\,m) with 12 participants.\footnote{Accuracy denotes the average angular offset between the calculated gaze ray and an imaginary gaze ray projected from its origin onto the target. Given a target, precision denotes the standard deviation of the calculated PORs.} At each distance, seven virtual targets were arranged in a cross-format at known spatial coordinates parallel to the y-z-plane. Targets were placed 0\textdegree{}, 10\textdegree{}, and 15\textdegree{} from the center. 
Targets were presented in random order for three seconds each. Participants were asked to fixate the stimulus and press a button on a PC mouse once they perceived a color change, which occurred after two seconds. This was employed to keep the attention of the participants on the target (\cf{ \cite{Feit2017TowardDesign}}). We started recording PORs after one second for one second ($\sim30$\,PORs), thus stopping when the color change happened. The average accuracy was 1.67\textdegree{} (0.5\,m), 0.51\textdegree{} (1\,m), and 0.47\textdegree{} (2\,m), with corresponding average precision over all targets of 1.21 (0.5\,m), 0.30 (1\,m), and 0.26 (2\,m). 
Our system estimates PORs by calculating where a user’s gaze ray intersects the spatial surroundings. To achieve this, we modeled the spatial surroundings using Unity, so that the virtual model of the real world overlays the physical world. Hence, the dimensions and locations of AOIs correspond to the dimensions and locations of the physical objects of interest. Using this approach, one can determine in which AOI any POR is located by simply assessing to which AOI the coordinates of that POR belong. 

\subsection{Implemented Eye Gaze Metrics}
To infer quantitative eye gaze metrics, we first identify fixations and saccades from PORs. Here, we use a dispersion-based algorithm designed to detect fixations in 3D, which is described more thoroughly in \cite{Weber2018Gaze3DFix:Volume}. This approach makes use of ellipsoidal bounding volumes whose size depends on the distance between the user and the fixation point. For each POR, we identify a fixation by checking whether the point lies within the ellipsoidal bounding volume. Now, given a set of successive PORs that have been classified as a fixation, we attribute that fixation to an AOI if any of the PORs of that set lie within the AOI. In other words, we attribute a fixation consisting of a set of PORs to an AOI if the two intersect. Based on the identified fixations, the system calculates the following gaze metrics:
\begin{enumerate}[leftmargin=*]
   \setlength\itemsep{-0.1em}
    \item \emph{Number of fixations}: For each AOI, we count the the number of fixations.
    \item \emph{Dwell duration}: For each AOI, we define dwell duration as the difference in seconds between the first fixation after entering the AOI and the last fixation before exiting the same AOI.
    \item \emph{Inter-POR distance of scanpath}: A series of PORs constitutes a scanpath in three-dimensional space. We define the inter-POR distance of such a scanpath as the average spatial distance between successive PORs.
    \item \emph{Angular distance}: We determine the visual angle $\theta$ at time $t$ between two successive gaze rays. Specifically, it is calculated through the dot product of two three-dimensional gaze direction vectors $g_{t}$ and $g_{t-1}$, \ie,
    $
        \theta_t = \mathrm{arccos}\left(\frac{\langle g_{t-1}, g_{t}\rangle}{\langle|g_{t-1}|, |g_{t}|\rangle}\right) .
    $
    \item \emph{TTFF}: TTFF on AOIs are calculated as the the time difference between starting the task and fixating the AOI for the first time.
\end{enumerate}

\section{User Study}
We explore gaze patterns under the guidance of different AR-based visual cues through a user study with $N=12$ participants. Specifically, we assessed $C=10$ experimental conditions, \ie, 2 baseline conditions and 8 AR-based visual cues.

\subsection{Experimental Setup and Task}
\begin{figure}[h]
\centering
  \includegraphics[width=0.999\columnwidth]{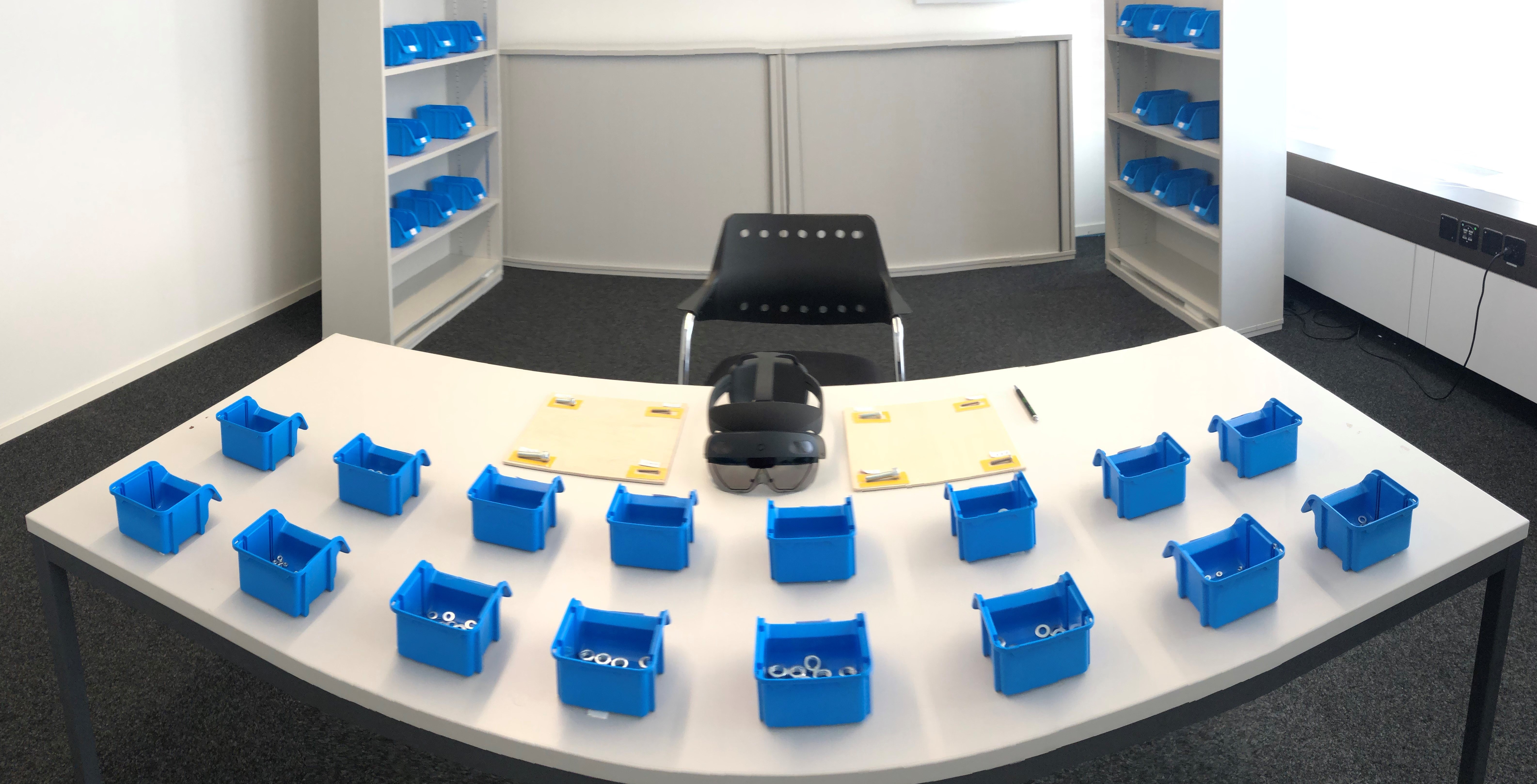}
  \caption{180° view of the experimental setup. Screws are located on the workpiece carrier, nuts and washers in the picking bins.}
  \label{fig:setup}
\end{figure}

We implemented a simulated assembly task similar industrial ones \cite{Guo2015OrderDisplays}. Participants assembled parts on a workpiece carrier, specifically washers and nuts that had to be turned on different screws. The screws were located on a central board in front of the participant. The corresponding nuts and washers were distributed across (a)~picking bins on a table and (b)~picking bins located in shelves at the end of the room (see Fig.~\ref{fig:setup}).

We used different assembly parts as follows. Screws varied by size (\ie, 4, 6, 8, and 10\,mm) and by type (\ie, \textquote{A} and \textquote{B}). This gives 24 distinct assembly parts (\ie, 8 screws, and, analogously, 8 nuts and 8 washers). As described above, nuts and washers were placed in picking bins either within the FOV (on the table) or outside the FOV (on the shelves, see Fig.~\ref{fig:setup}). Picking bins and screws were labeled according to the assembly parts by a three-letter code, referring to the size, type, and assembly part. For instance, the picking bin of an 8\,mm type~\textquote{A} nut was labeled by {\textquote{8-A-N}}.


In our experimental task, participants were asked to assemble $M$ target screws with the corresponding nuts and washers. During this, participants were supported by different cues as defined by the experiment condition. Depending on condition, the number of target screws to be assembled, $M$, was varied (\ie; $M = 2$ or $M = 3$; see Sec.~\ref{exp_design}). Participants were allowed to choose the order of assembly. That is, participants were free to attach the nut on the screw first or the washer, if both were highlighted simultaneously in the respective experimental conditions. Further, no time limit was imposed.

\begin{figure}[!b]
    \centering
    \sbox0{\includegraphics[trim={3.5cm 0 4cm 0}, width=0.92\linewidth]{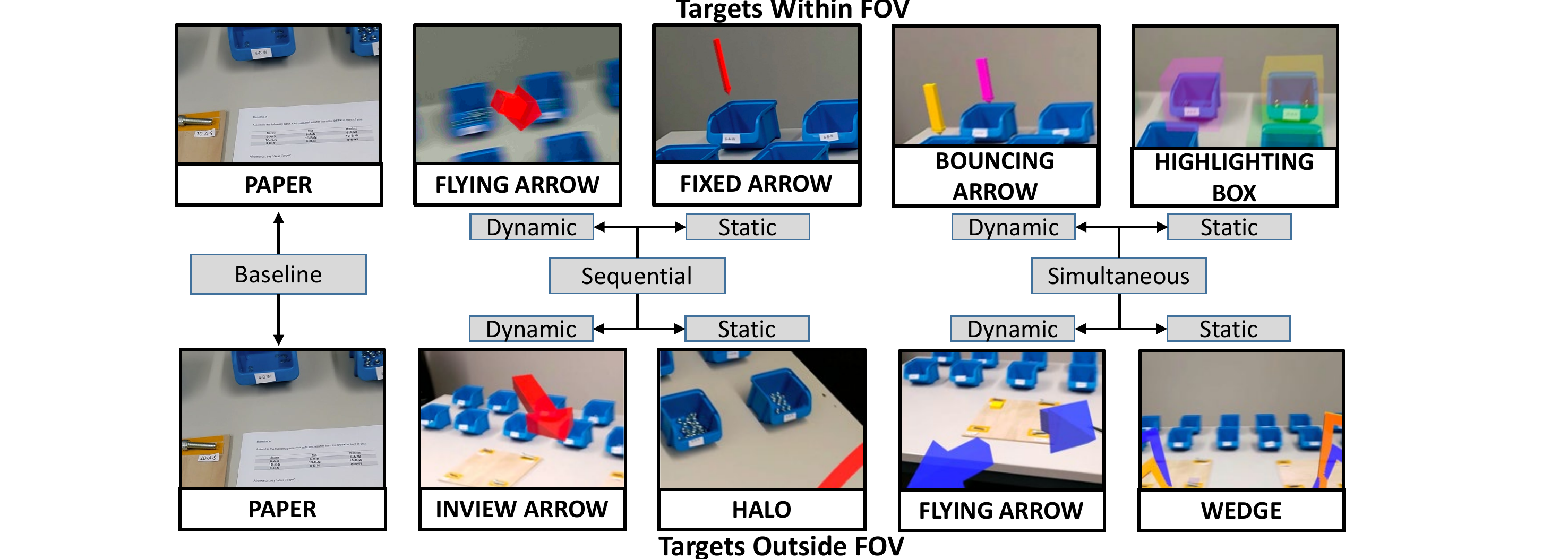}}
    \begin{minipage}{\wd0}
    \usebox0
    \caption{Ten experimental conditions, which are grouped by within FOV (top) and outside FOV (bottom).}
    \label{fig:conditions}
    \end{minipage}
\end{figure}

\subsection{Experimental Design}
\label{exp_design}

We conducted a within-subject user study with $C=10$ conditions (2 baselines and 8 AR-based visual cues). The conditions were randomly assigned across $N=12$ participants. The conditions represent different cues that were used to show the targets in the experimental task. In the baseline conditions, participants received guidance through a piece of paper that stated the labels of the target screws and the labels of the corresponding nuts and washers. For all other conditions, participants were shown one or multiple AR-based visual cues. The complete list of conditions is given in Fig.~\ref{fig:conditions}.

The visual cues differed along various dimensions: (a)~whether user attention was guided to targets within or outside the FOV; (b)~whether user attention was guided by a single visual cue at a time or by multiple visual cues displayed simultaneously; and (c)~whether the cue appeared static or dynamic. Depending on the dimension of a visual cue, the experiment task was adapted as follows:
\begin{itemize}[leftmargin=*]
\setlength\itemsep{-0.2em}
\item \emph{Within vs. outside FOV}. In order to guide user attention to targets within the FOV, all target picking bins were located on the table in front of the participant and thus within the user's FOV (see Fig.~\ref{fig:setup}). In contrast, for guidance towards targets outside the FOV, picking bins were located on the two shelves outside of the FOV. 
\item \emph{Sequential vs. simultaneous}. Sequential cues displayed only a single visual marker at a time for the target picking bins. Simultaneous visual cues showed multiple visual markers at the same time. In the latter case, visual markers were differentiated by coloring them according to the different targets. As described above, participants were free to attend the highlighted picking bins in any order and without time restrictions.
\item \emph{Static vs. dynamic}. Static cues had no or only limited movement in the AR environment, while dynamic cues had some form of movement towards the target. 
\end{itemize}

Each visual cue was accompanied by a small semi-transparent highlighting box around the target screw (or screws). This highlighting box was shown in addition to the main visual cue in order to help participants identify the target screw. As a default, participants were asked to assemble $M=3$  target screws (selected at random). An exception was made for two visual cues (simultaneously displayed cues with targets outside FOV). Here, the number of target screws was set to $M=2$ in order to reduce visual clutter within the HMD. 

With the selection of visual cues (Fig.~\ref{fig:conditions}), we cover a range of common visual cues that have been found to be effective regarding task completion time or number of task errors (\eg, see \cite{Gruenefeld2017VisualizingReality, Renner2017AttentionSystems,Gruenefeld2018FlyingARrow:Devices}). We discarded guidance cues that were ill-suited for HoloLens, such as the \cue{attention funnel} \cite{Biocca2006}, because it is difficult to follow in small FOVs \cite{Renner2017AttentionSystems} and is limited to a single off-screen object at a time \cite{Gruenefeld2017EyeSee360:Reality}. In implementing the cues, we kept their visual appearance as close as possible to the respective original presentation.

\subsection{Procedure and Measurements}

Upon arrival, each participant received an information sheet summarizing the goals, methods, and compensation of the user study. After having time to ask questions, participants signed a consent form and filled out a demographics questionnaire. Participants were compensated with the equivalent of USD~20.

Participants first calibrated the HMD to their eyes for optimal hologram appearance and stable eye tracking. Before starting the experimental task, participants were introduced to the HMD in a training round in which the setting, task, and HMD controls were explained by an experimenter. Afterwards, each participant performed the experimental task for all $C=10$ conditions. Prior to conducting the user study, we obtained ethics approval from the Ethics Committee of ETH Zurich.

\subsection{Participants and Statistical Testing}

We recruited 12 participants (6 female, 6 male) aged between 23 and 35 ($M = 26.92$; $\mathit{SD} = 4.14$). No participant had color vision impairment or other binocular vision disorders. All participants had normal or corrected vision. Five participants reported experience with VR technology and one with AR technology.

We used a repeated measures ANOVA. In case the assumption of sphericity was violated, the degrees of freedom were adjusted using Greenhouse–Geisser correction. In case of violated ANOVA test assumptions, we used Friedman tests together with Scheff\`{e}'s method for multiple comparisons \cite{Scheffe1953AVariance}. This is a conservative method for pairwise comparisons and any number of non-pairwise comparisons of group means \cite{Lee2018WhatTest}. We report mean values ($M$) and median values, depending on the underlying statistical test.

\section{Results}
\subsection{Gaze Distribution}
\label{sec_gaze_distr}

We compare how visual cues guide user attention and thus affect gaze distribution based on the following AOIs: 
\begin{enumerate}[leftmargin=*]
    \setlength\itemsep{-0.1em}
    \item \textquote{\emph{On targets}} refers to all \emph{highlighted} targets. It includes all picking bins and assembly parts that are relevant for solving the current task (independent of whether they are within or outside the FOV) and that are marked by the visual cue. 
    \item \textquote{\emph{On potential targets}} refers to all \emph{candidate} targets. It includes all picking bins and assembly parts that have not been assigned to target objects.
\end{enumerate}

\begin{figure}[h]
\centering
\begin{subfigure}[t]{0.80\linewidth}
   \includegraphics[trim={0cm 0cm 0cm 0cm}, width=0.90\linewidth]{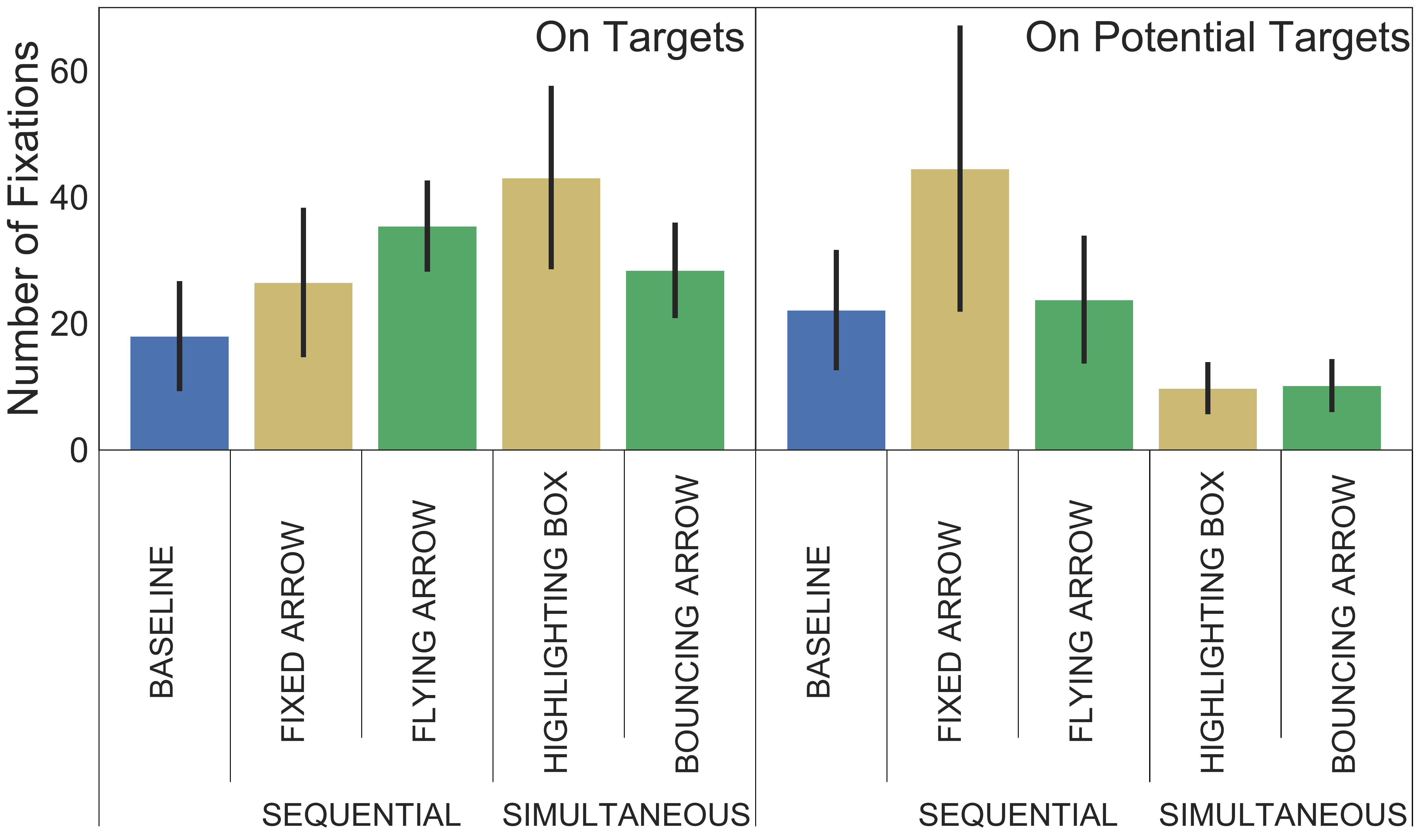}
   \caption{Conditions: Targets within FOV.}
   \label{fig:distr_sub_1} 
\end{subfigure} \\
\begin{subfigure}[t]{0.80\linewidth}
   \includegraphics[trim={0cm 0cm 0cm 0cm}, width=0.90\linewidth]{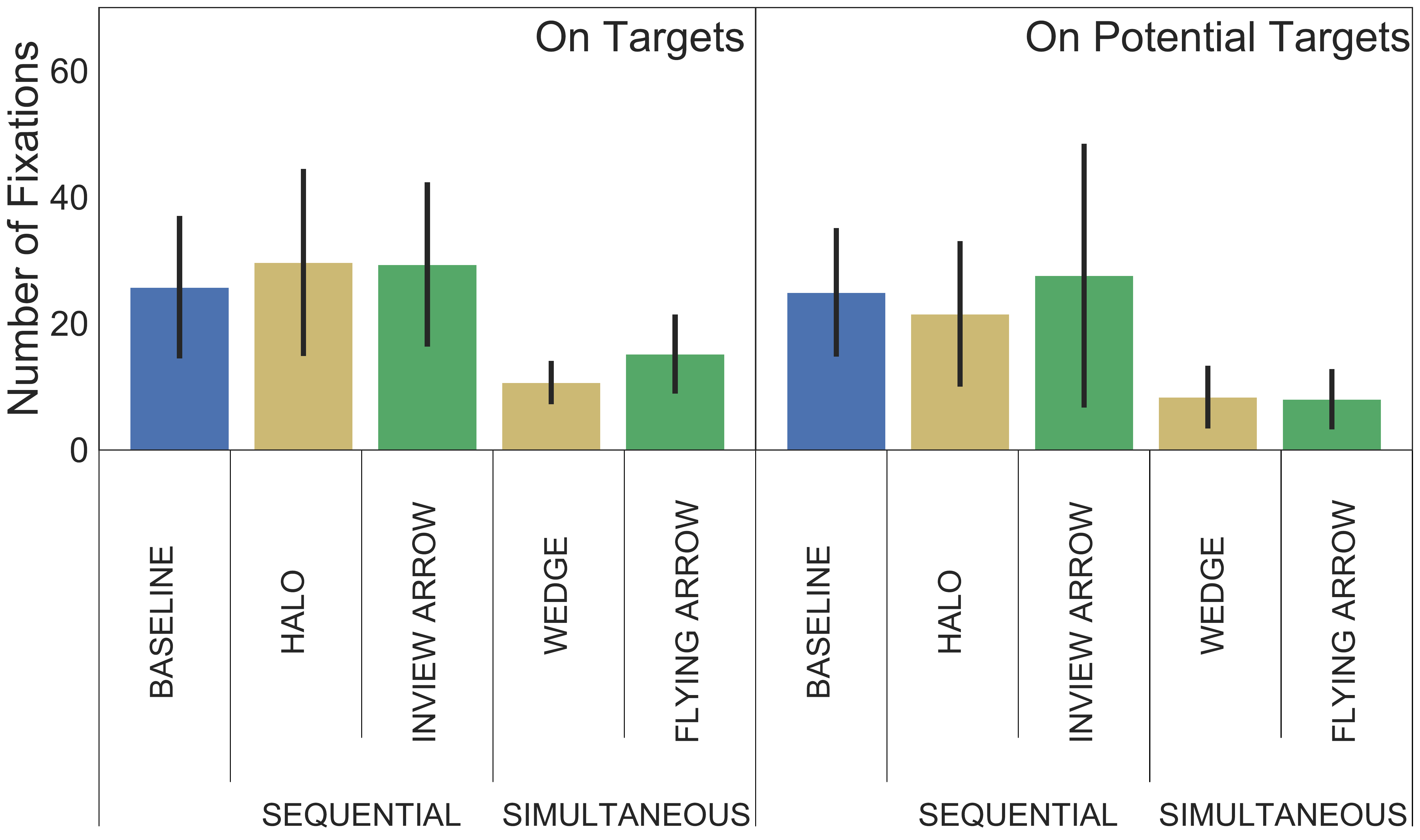}
   \caption{Conditions: Targets outside FOV.}
   \label{fig:distr_sub_2}
\end{subfigure}
\caption{Eye fixations on targets and potential targets across different experimental conditions. Whiskers show standard deviations.}
\label{fig:gaze_distribution}
\end{figure}

Results for gaze distributions (\ie, eye fixations per AOI) are shown in Fig.~\ref{fig:gaze_distribution}.
For the conditions with \textbf{targets within FOV}, visual cues had a profound effect on the gaze distribution for \textquote{on targets} (Fig.~\ref{fig:distr_sub_1},~left): The number of fixations on targets was significantly affected by the presence of a visual cue ($F(4,44) = 12.5, p<0.001$). All means were higher than that of the \cue{Baseline}. In other words, all AR-based cues captured more eye fixations on targets than the \cue{Baseline} (non-pairwise comparison, $p<0.01$). The largest number of eye fixations was attributed to the \cue{Highlighting Boxes} cue ($M=43.08, SD =14.51$), which generated more eye fixations than all other conditions (all pairwise tests with $p<0.05$, except for the \cue{Flying Arrow}). For \textquote{on potential targets} (Fig.~\ref{fig:distr_sub_1},~right), the number of fixations also differed significantly between conditions ($F(1.61, 17.66)=18.05, p<0.001$). More specifically,  simultaneous cues ($M=12.83, SD=4.30$) attracted fewer eye fixations than sequential cues ($M=37.38, SD=20.82, p<0.001$). Conversely, the \cue{Fixed Arrow} had the largest number of eye fixations on potential targets ($M=47.25, SD=24.10$) with significant differences (pairwise) to all other visual cues ($p<0.05$).

For the condition with \textbf{targets outside the FOV} and fixations \textquote{on targets} (Fig.~\ref{fig:distr_sub_2}, left), group means differed significantly across cues ($F(4,44)=11.04$; $p<0.001$). Moreover, non-pairwise comparisons of sequential ($M=29.50$; $\mathit{SD}=13.62$) and simultaneous ($M=12.92$; $\mathit{SD}=5.44$) visual cues showed significant differences ($p<0.001$). In this regard, both baseline and sequential cues captured eye gaze to a similar extent. However, simultaneous cues generated a considerable lower number of eye fixations on targets. Similar patterns are observed for \textquote{on potential targets} (Fig.~\ref{fig:distr_sub_2}, right). The average number of fixations on potential targets were significantly different among the experimental conditions ($F(4,44)=11.15, p<0.001$). Likewise, comparing sequential ($M=27.71,SD=17.20$) and simultaneous ($M=10.33, SD=5.83$) visual cues yielded significant differences ($p<0.001$). 

\subsection{Gaze Duration}

\begin{figure}[!h]
\centering
\begin{subfigure}[t]{0.80\linewidth}
   \includegraphics[width=0.90\linewidth]{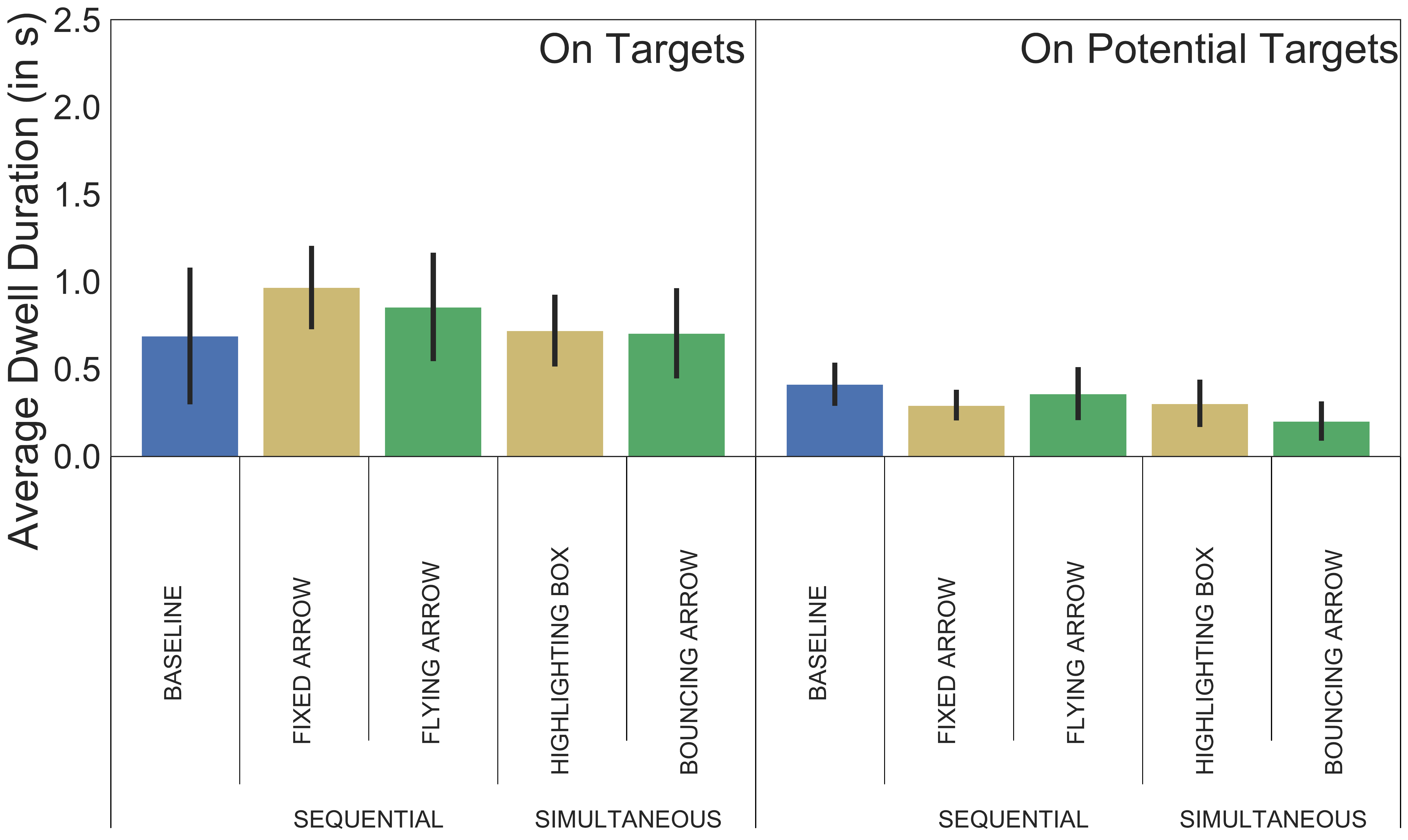}
   \caption{Conditions: Targets within FOV.}
   \label{fig:dwell_time_sub_1} 
\end{subfigure} \\
\begin{subfigure}[t]{0.80\linewidth}
   \includegraphics[width=0.90\linewidth]{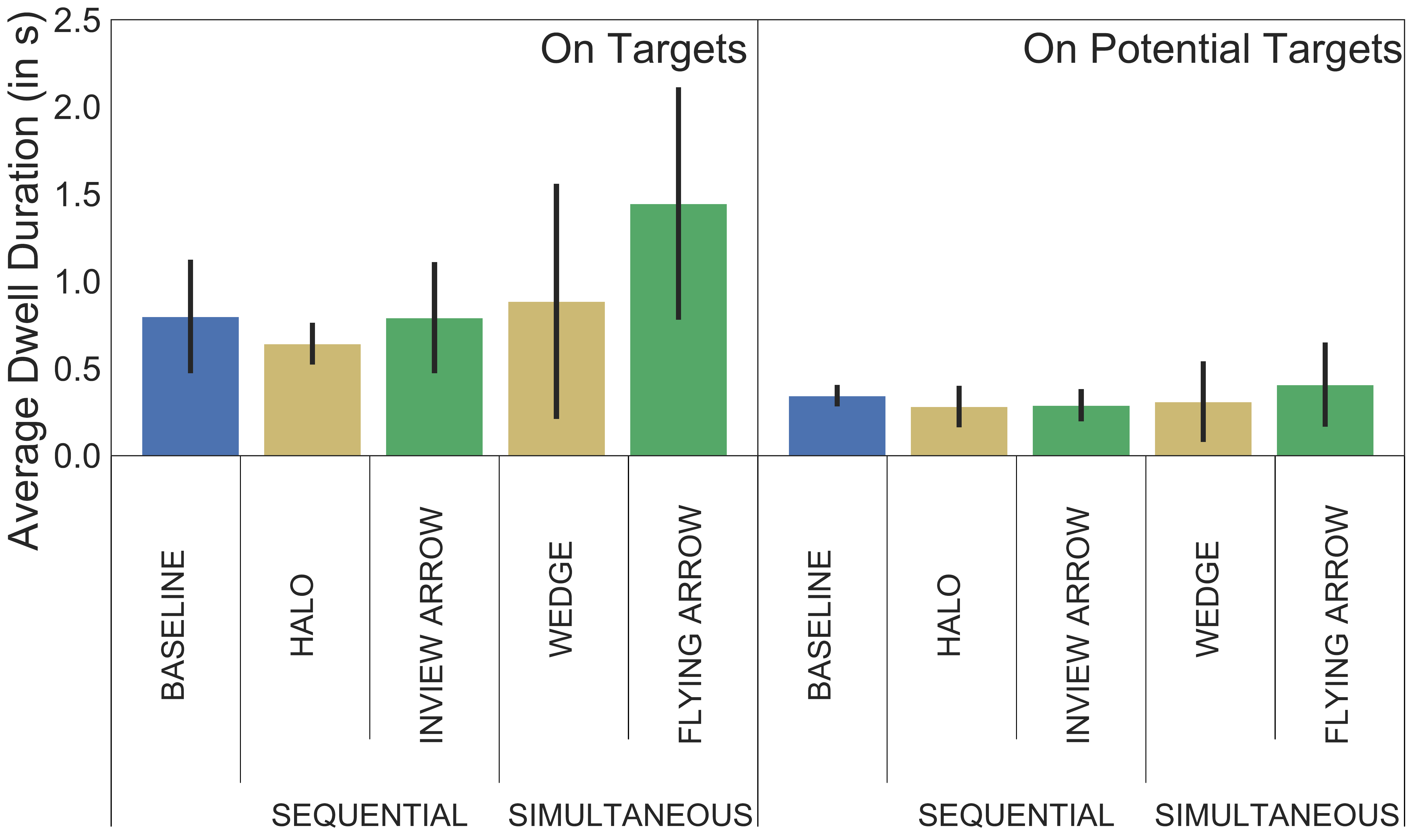}
   \caption{Conditions: Targets outside FOV.}
   \label{fig:dwell_time_sub_2}
\end{subfigure}
\caption{Dwell times on targets and on potential targets across different experimental conditions.  Whiskers show standard deviations.}
\label{fig:avg_dwell_time}
\end{figure}

We report gaze duration by comparing average dwell times among different conditions (Fig.~\ref{fig:avg_dwell_time}). Here, we draw upon the same AOIs as before, namely \textquote{on targets} and \textquote{on potential targets}.

For conditions with \textbf{targets within FOV} (Fig.~\ref{fig:dwell_time_sub_1}), the average dwell duration for \textquote{on targets} differed significantly ($\chi^2(4)=12.00$; $p<0.05$). However, comparing sequential with simultaneous visual cues and static with dynamic ones did not lead to significant differences. For \textquote{on potential targets}, the mean dwell duration varied significantly across conditions ($\chi^2(4)=23.73$; $p<0.001$). Likewise, comparing sequential vs. simultaneous cues and static vs. dynamic cues did not lead to significant differences. We further conducted pairwise comparisons: The simultaneous dynamic visual cue \textsc{Bouncing Arrow} (median $=0.18$) led to a significant difference over the baseline condition (median $=0.39$; $p<0.01$), as did the \textsc{Fixed Arrow} (median $= 0.28, p<0.05$). 

For conditions with \textbf{targets outside FOV} (Fig.~\ref{fig:dwell_time_sub_2}), average dwell duration for \textquote{on targets} varied significantly ($\chi^2(4)=13.53$; $p<0.01$). A non-pairwise comparison of sequential visual cues (median $= 0.72$) and simultaneous ones (median $=1.16$) showed significant differences ($p<0.05$). The visual cue \textsc{Flying Arrow} (simultaneous dynamic) generated the longest average dwell duration (median $=1.45$). However, a pairwise comparison to all conditions was not significant, except for the simultaneous static visual cue (\cue{Wedge}, $p<0.05$). We did not find any significant deviations in group means for a dwell duration \textquote{on potential targets}.

\subsection{Inter-POR Distance of Scanpath}
\begin{figure}
\centering
\begin{subfigure}[t]{0.49\linewidth}
   \includegraphics[width=0.9\linewidth]{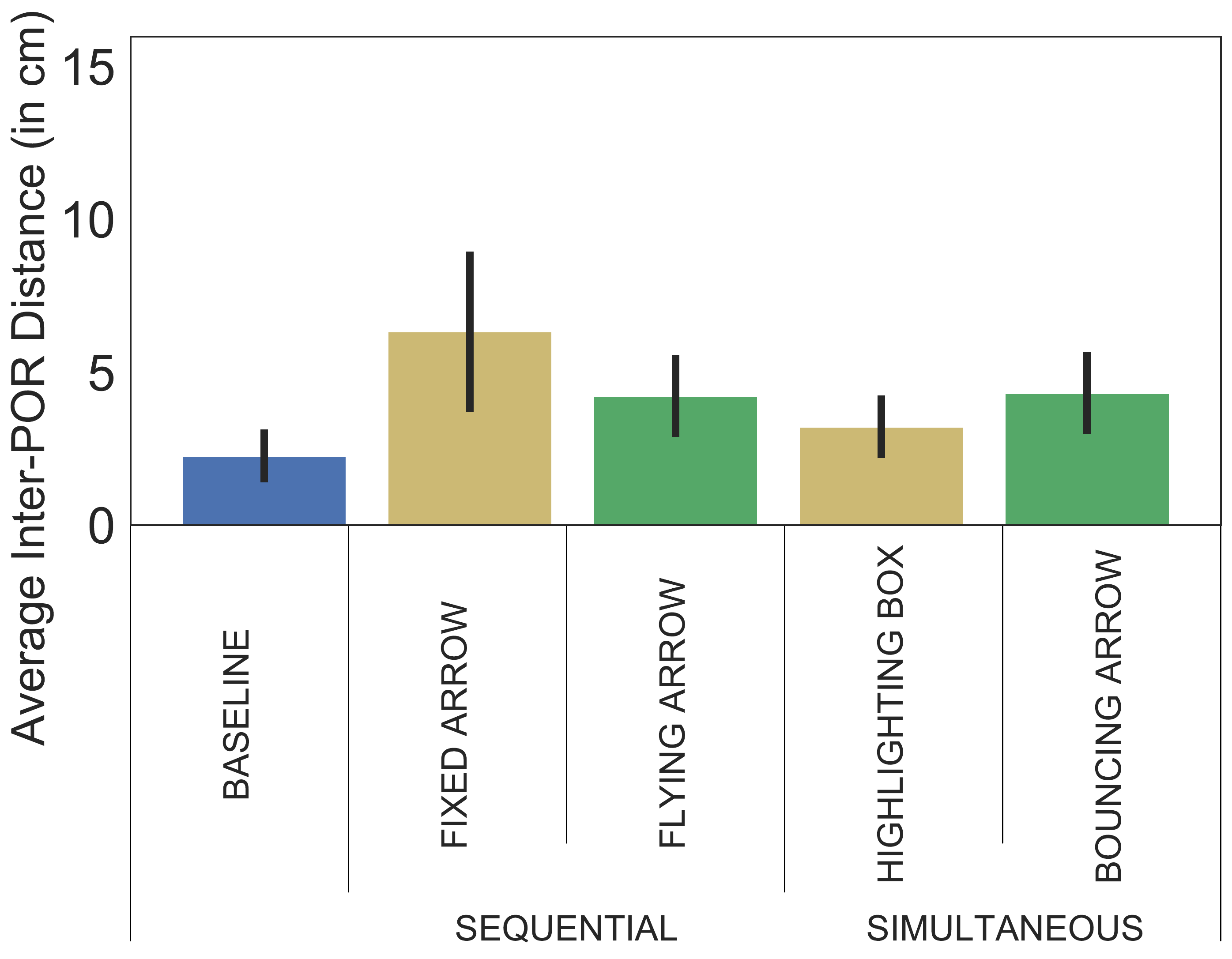}
   \caption{Conditions: Targets within FOV.}
   \label{fig:scanpath_iv} 
\end{subfigure} \hfill
\begin{subfigure}[t]{0.49\linewidth}
   \includegraphics[width=0.9\linewidth]{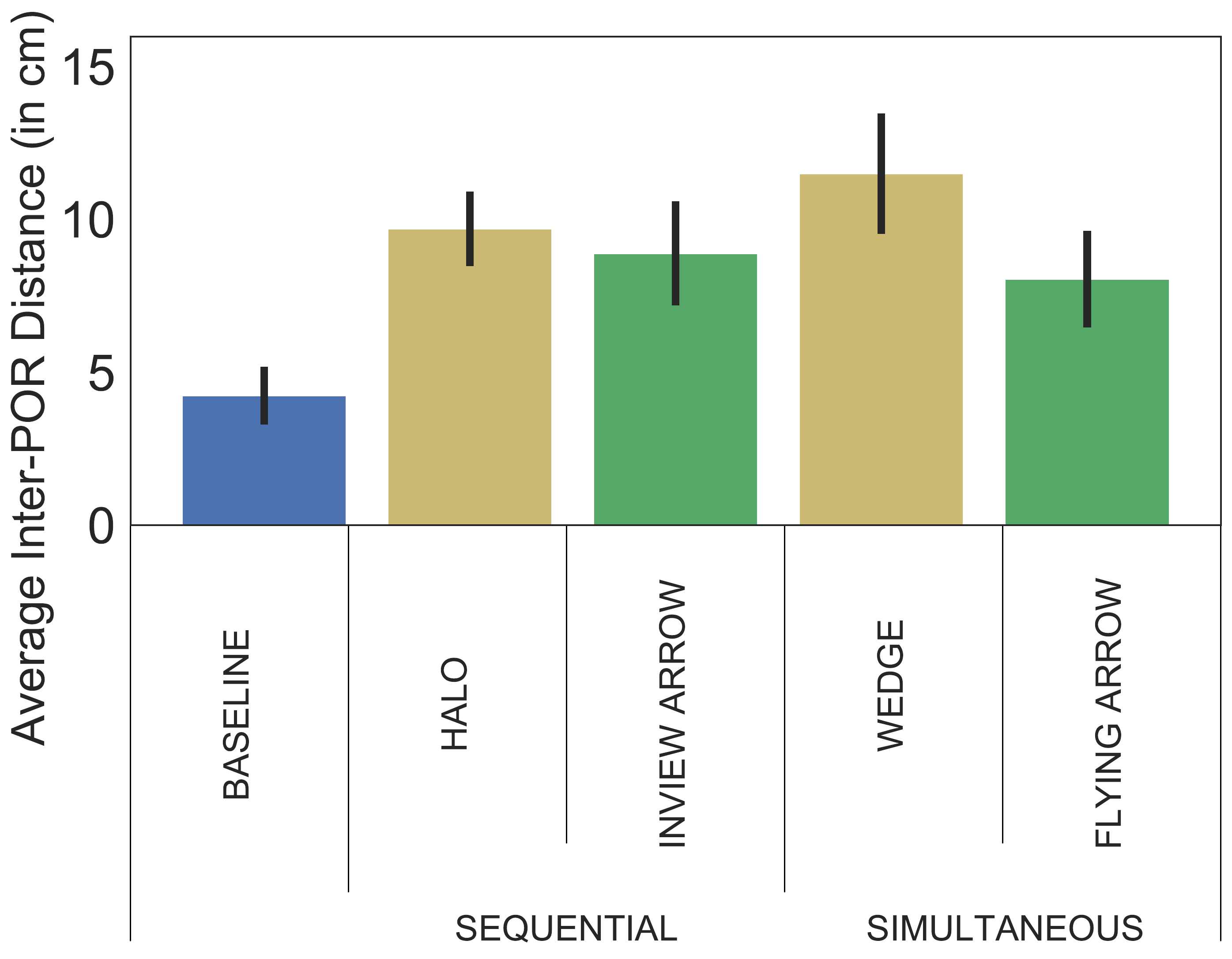}
   \caption{Conditions: Targets outside FOV.}
   \label{fig:scanpath_ov}
\end{subfigure}
\caption{Inter-POR distances across different experimental conditions. Whiskers show standard deviations.}
\label{fig:scanpath}
\end{figure}

We report the following results for the average distance between consecutive PORs (see Fig.~\ref{fig:scanpath}). 
For conditions with \textbf{targets within FOV}, group means were significantly different ($F(1.78, 19.54)=14.41$; $p<0.001$). Here, a non-pairwise comparison established that visual cues resulted in a higher average inter-POR distance ($M = 4.53$; $\mathit{SD}=2.00$) as compared to the \textsc{Baseline} ($M = 2.27$; $\mathit{SD}=0.86$; $p<0.01$). Furthermore, the inter-POR distance was higher for sequential cues ($M=5.29$; $\mathit{SD}=2.30$) as compared to simultaneous cues ($M=3.77, \mathit{SD}=1.30, p<0.05$).

For conditions with \textbf{targets outside FOV}, group means significantly differed ($F(4,44)=48.36$; $p<0.001$). Based on a non-pairwise comparison, we found that the average inter-POR distance was higher for cues ($M = 9.54$; $\mathit{SD}=2.04$) as compared to the \cue{Baseline} ($M =4.24$; $\mathit{SD}=0.94$; $p<0.001$). Furthermore, dynamic cues showed a shorter inter-POR distance ($M=8.47$; $\mathit{SD}=1.66$) as compared to static ones ($M=10.60$; $\mathit{SD}=1.85$; $p<0.001$).

\subsection{Time to First Fixations}
\begin{figure}[h]
\centering
\begin{subfigure}[t]{0.9\linewidth}
   \includegraphics[width=0.8\linewidth]{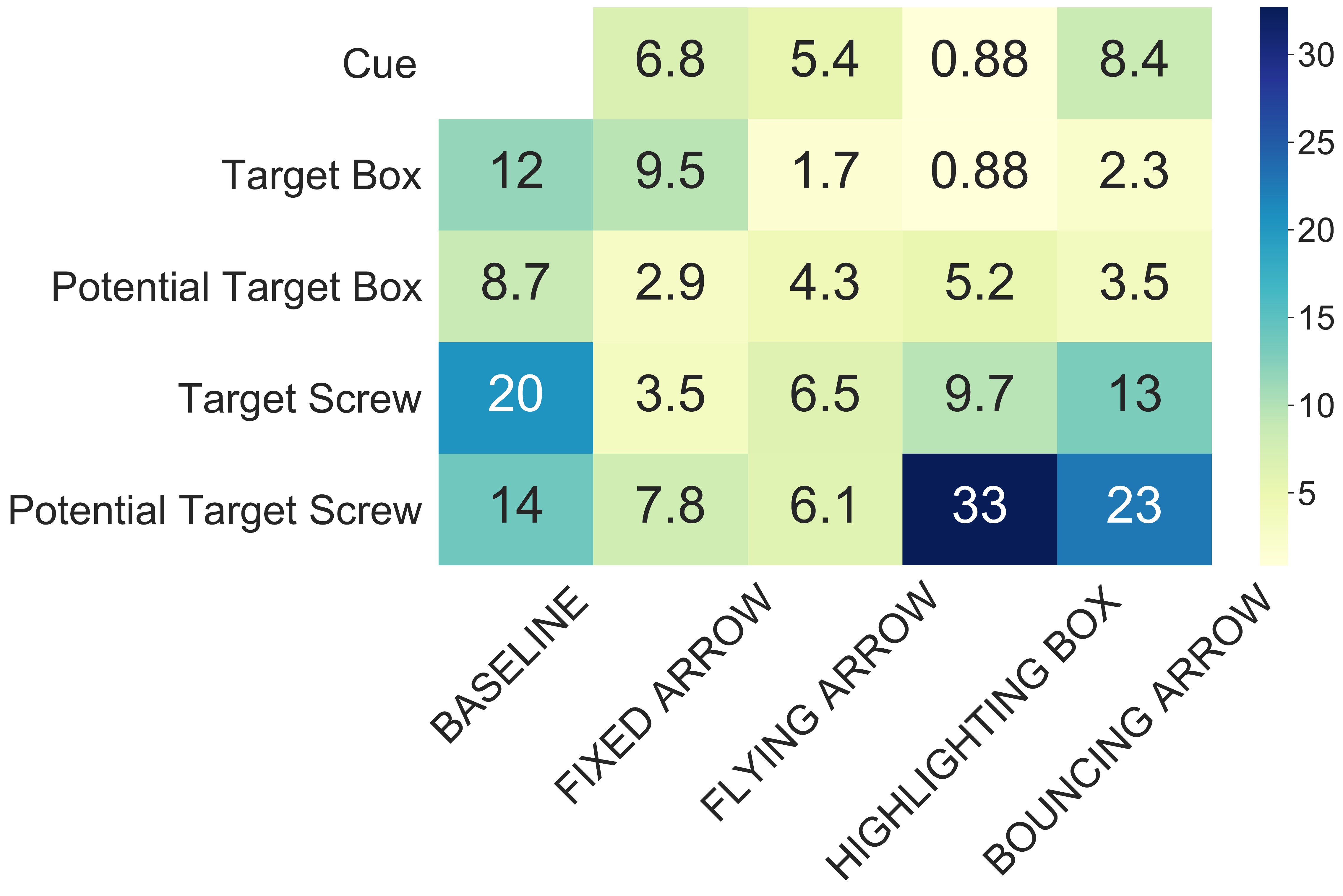}
   \caption{Conditions: Targets within FOV.}
   \label{fig:ttff_sub_1} 
\end{subfigure} \\
\begin{subfigure}[t]{0.9\linewidth}
   \includegraphics[width=0.8\linewidth]{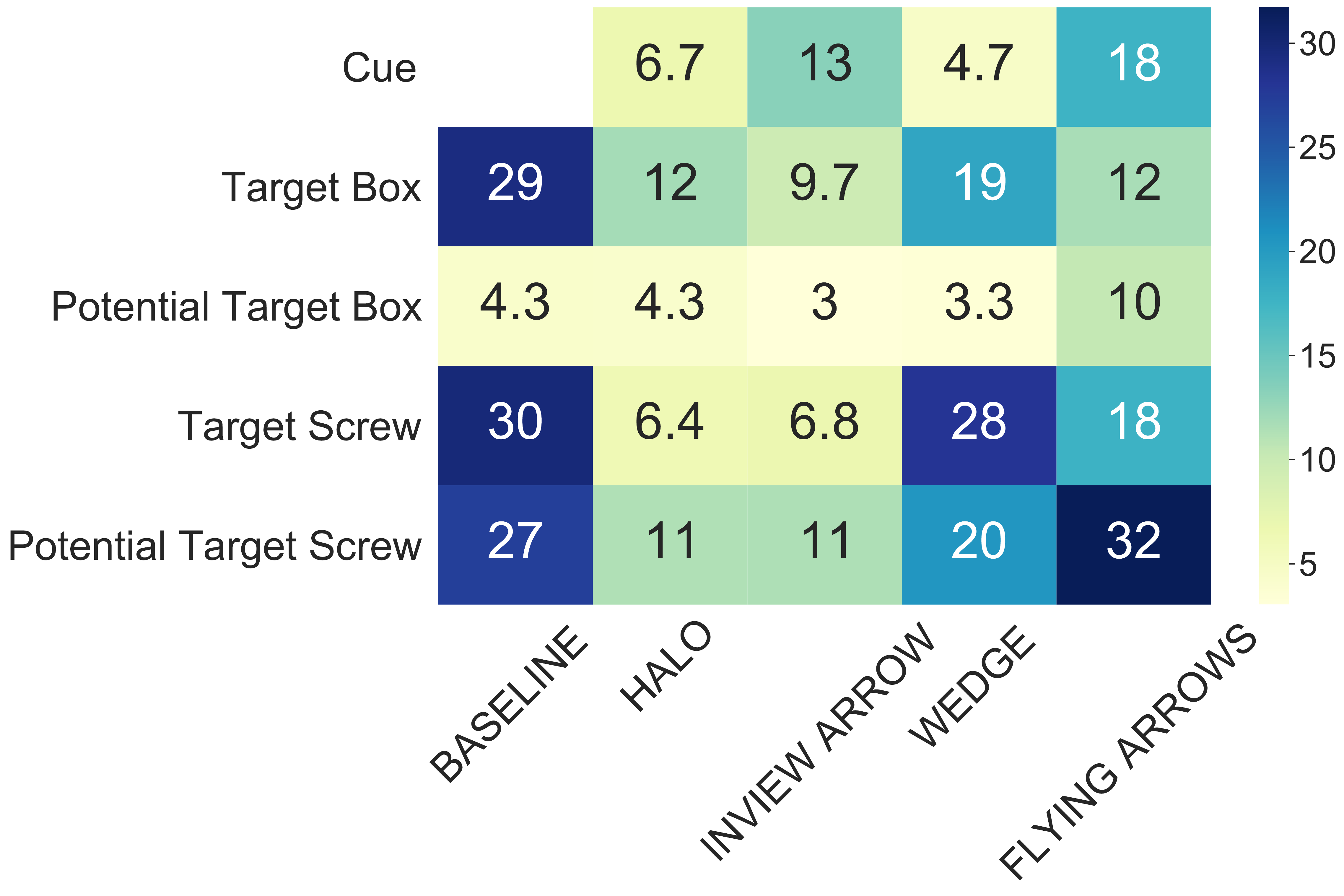}
   \caption{Conditions: Targets outside FOV.}
   \label{fig:ttff_sub_2}
\end{subfigure}
\caption{TTFF (seconds) across AOIs and experimental conditions.}
\label{fig:ttf_heatmap}
\end{figure}

We report average TTFFs (in seconds) for the following AOIs: (1)~\emph{Cue} represents the AR-based visual cue (\ie, implemented for all conditions except for the two baselines). (2)~\emph{Target boxes} and (3)~\emph{target screws} refer to the picking bins and screws, respectively. Both have been highlighted for the participants to solve the current task at hand.  (4)~\emph{Potential target boxes} and (5)~\emph{potential target screws} comprise of any other picking bins and screws, respectively, that are not assigned to the targets and thus not highlighted by any visual marker. Fig.~\ref{fig:ttf_heatmap} reveals large differences in TTFF among the visual cues and AOIs. We report the highest $p$-value that applies to the mentioned conditions.

For conditions with \textbf{targets within FOV}, TTFFs differed significantly among visual cues for each AOI ($p<0.001$). The only exception are the potential target boxes ($p=0.16$). For both types of targets (box and screw), TTFF was lower when visual cues were shown as compared to the \cue{Baseline} ($p<0.05$). Moreover, sequential cues had a significantly lower TTFF on potential screw targets as opposed to simultaneous ones ($p<0.001$).

For the conditions with \textbf{targets outside FOV}, we found no significant differences in TTFFs regarding cues and potential targets screws. All other AOIs had significantly different TTFFs across conditions ($p<0.01$). Looking at target boxes, TTFF was lower when visual cues were shown as compared to the \cue{Baseline} ($p<0.01$). Regarding target screws, sequential visual cues resulted in a lower TTFF than simultaneous ones ($p<0.05$).

\section{Discussion}
Our results show that visual attention was strongly affected by the presence of cues. This finding is in line with works reviewed above that highlight the potential for AR-based visual cues to shift fixations towards objects of interest (\eg, \cite{Burova2020UtilizingMaintenance}). Across all visual cues and positions of targets (within FOV and outside FOV), users fixated target objects more quickly when being guided by visual cues, which has been identified as an important criterion for successful attention guidance \cite{Renner2017AttentionSystems}. Additionally, visual cues increased the average inter-POR distance of participants' scanpaths. Larger distances between successive PORs suggest that visual cues are regarded as meaningful, as the cues guide user attention to the desired targets more directly and with less interim PORs \cite{Goldberg1999ComputerConstructs}. Similarly, scanpath length has been considered as an indicator for efficiency or productivity of interfaces \cite{Goldberg1999ComputerConstructs, Poole2005EyeProspects}. Overall, these findings not only corroborate existing studies on the effectiveness of AR-based visual cues for attention guidance (\eg, \cite{ Renner2017AttentionSystems, Jeffri2020GuidelinesAssembly}), but also provide first insights into their underlying eye gaze mechanisms. It is worth mentioning, however, that some effects were more pronounced depending on whether the target object was within or outside the FOV.

\subsection{The Effect of Simultaneous Cues on Gaze Behavior}
Simultaneous cues were successful in shifting fixations away from non-target objects, which can be seen as an indicator for unhindered or efficient search \cite{Goldberg1999ComputerConstructs}. For cues with targets within the FOV, we observed a substantial number of fixations on the target boxes. For simultaneous cues with targets outside the FOV, however, the opposite was observed (fewer fixations on target boxes). One potential reason might be that participants fixated these cues instead of the target boxes. Irrespective of this, we summarize that simultaneously presented cues were more successful in shifting attention away from non-target objects than sequential ones. This could, possibly, arise from the difference between parallel and serial search, which plays a central role in, for example, FIT \cite{Treisman.A1980AAttention}. 

Analyzing other gaze metrics like TTFF, we found some cues to affect gaze patterns very strongly and others to have a small effect. In particular, for cues with targets within the FOV, the simultaneous \cue{highlighting boxes} led to many quick fixations on targets. One cause might be that this cue is an exogenous one, attracting bottom-up (stimulus-driven) attention \cite{Jeffri2020GuidelinesAssembly}. In contrast, cues like the \cue{flying arrow} can be seen as endogeneous, since they point towards the real target, thus requiring a process under attentional control \cite{Jeffri2020GuidelinesAssembly}. While this is consistent with other studies \cite{Theeuwes2010Top-downSelection}, more empirical evidence is needed to investigate this question. For cues with targets outside the FOV, we found \cue{Wedge} to not only lead to the fewest fixations on targets, but also their slowest TTFF. Due to multiple in-view elements, this cue might lead to visual clutter (especially for small FOVs \cite{Gruenefeld2018FlyingARrow:Devices}), which is linked to increased difficulty of visual search  \cite{Rosenholtz2007MeasuringClutter}, thereby explaining increased TTFF on targets. 

\subsection{The Effect of Cue Motion on Gaze Behavior}
When presented with dynamic cues, users fixated the target boxes more quickly than the visual cue. The opposite pattern occurred across all static conditions, in which users first fixated the visual cue and then the target boxes. This result suggests that users did not need to look at the dynamic cue to extract the required information from it, which is plausible given that visual information is not only gathered at the point of eye fixation, but also in the visual periphery \cite{eckstein2004decoupling}. 

Our results further show that cues with more motion (\eg, \cue{Flying Arrow}) revealed pronounced effects on gaze behavior, as compared to cues with limited motion, such as the rotating \cue{Inview Arrow}. 
This might be explained by differences in visual saliency among the cues. Visual saliency is is a composite metric of many low-level bottom-up features, including both size and motion, and it is generally linked to attention capture \cite{Itti2001ComputationalAttention}.

\subsection{Conclusion and Future Research}
\label{future_work}
In this paper, we explored the effects of visual guidance cues on eye gaze patterns in AR through a user study with $12$ participants. Different visual cues were displayed via an AR HMD while eye movements were tracked simultaneously. Although we found our system to measure with high accuracy and precision, our findings are limited by the maximum sample frequency of 30\,Hz. Thus, future studies could strive to utilize low-latency eye tracking systems for AR with high sampling frequencies. This could, for example, be beneficial for assessing additional eye gaze patterns like smooth pursuits. Further, we focused on a variety of conventional gaze-related metrics. However, a multitude of complementing metrics might be informative, such as transition probabilities among AOIs.
\bibliographystyle{abbrv}
\bibliography{eye_gaze_bib}
\end{document}